# Influence of Topology on the Ultrafast Carrier Dynamics in MoTe$_2$

Sara L. Zachritz[1], Joohyung Park[1], Ayan Batyrkhanov[1], Leah L. Kelly[2], Dennis L. Nordlund[2], and Oliver L.A. Monti[1,3,*]

[1]*Department of Chemistry and Biochemistry, University of Arizona, 1306 East University Boulevard, Tucson, AZ 85721, USA*

[2]*SLAC National Accelerator Laboratory, Stanford Synchrotron Radiation Lightsource, 2575 Sand Hill Road, MS 99, Menlo Park, CA 94025, USA*

[3]*Department of Physics, University of Arizona, 1118 East Fourth Street, Tucson, AZ 85721, USA*

*Corresponding author: monti@arizona.edu

**Abstract:**

Transport properties of Weyl semimetals are intimately connected to the underlying band structure. The signature of Weyl semimetals are linear dispersing bands that touch, forming Weyl points that result in high charge carrier mobilities. The layered transition metal dichalcogenide MoTe$_2$, undergoes a temperature dependent phase transition that directly converts the trivial 1T' phase to the nontrivial T$_d$ phase, providing an opportunity to understand how the formation of a Weyl point manifests in the ultrafast carrier dynamics. In this study, we use resonant X-ray photoemission to monitor the element specific evolution of excited carriers 1T'-MoTe$_2$ and in the vicinity of the Weyl point in T$_d$-MoTe$_2$. We find that the delocalization time of 1T'-MoTe$_2$ is a factor 1.5 times faster than in T$_d$-MoTe$_2$. We argue that this is a result of the change in the density of states and screening length, to a higher carrier scattering rate in T$_d$-MoTe$_2$. Our study tracks the fate of carriers in MoTe$_2$ on sub-fs time-scales and with atomic site specificity.

**Introduction:**

The discovery of Weyl semimetals (WSMs) has added a new class of materials to the family of topologically non-trivial quantum matter.[1,2] WSMs are inversion-symmetry broken and host Weyl points, where linearly dispersing valence and conduction bands form electron and hole pockets that touch near the Fermi energy, $E_F$. The Hamiltonian in the vicinity of these points can be reduced to a realization of the relativistic Weyl equation. Weyl points are symmetry-protected such that low-energy excitations are seemingly unaffected by crystal imperfections. As a consequence, backscattering is suppressed, and WSMs support high charge carrier mobilities and extremely high magnetoresistance.[3–5] In this way, WSMs provide new opportunities for creating dissipationless electronic and spintronic devices.

Though the discovery of WSMs remains challenging, one of the most extensively investigated candidates is $T_d$-MoTe$_2$, recently confirmed as a type-II topological WSM.[6–11] Unique to $T_d$-MoTe$_2$ is that it is related to the topologically trivial 1T'-phase, connected by a temperature-dependent structural phase transition. 1T'-MoTe$_2$ has a monoclinic crystal structure (space group $P2_1/m$), with preserved inversion symmetry. Below ~240 K, the inversion symmetry breaks as 1T'-MoTe$_2$ turns into the topologically nontrivial $T_d$-phase with an orthorhombic crystal structure (space group $Pmn2_1$).[12,13]

To explore the potential of WSMs for enabling new modalities for ultrafast electronics and spintronics, a deeper understanding of carrier dynamics in these materials is needed. The readily accessible phases in MoTe$_2$ provide thus an unparalleled platform to compare the carrier dynamics in closely related topologically trivial and nontrivial materials. Indeed, the relaxation pathways of photoexcited carriers with momentum resolution in 1T'- and $T_d$-MoTe$_2$ have already been investigated using ultrafast pump-probe time- and angle-resolved photoemission spectroscopies.[14–17] However, the ultrafast dynamics accessed in these studies are not only limited to approximately the pulse duration, usually on the order of a few tens of femtoseconds, but also reflect the delocalized nature of optical excitations and therefore the resulting relaxation pathways. In this study, we use core-hole clock (CHC) spectroscopy, a resonant soft X-ray approach, to access timescales on the order of attoseconds (as) to a few fs. This is made possible by using an internal reference "clock" provided by the lifetime of the core hole.[18–22] Unlike typical pump-probe spectroscopies, CHC naturally provides element- and site-specificity due to the localized nature of the core electrons and the chemically specific nature of core level electrons. Additionally, selection rules for excitation of core electrons can even provide orbital specificity, thereby accessing select regions of the Brillouin zone. As a consequence, the localized nature of these core level excitations in both real and reciprocal space ultimately reveals ultrafast dynamics that are sensitive to changes in the specific chemical environment. CHC is thus ideally suited to investigate the electronic consequences of the structural and topological 1T'→$T_d$ phase transition that occurs in $MoTe_2$.

Here, we probe the evolution of excited states associated with tellurium orbital character in the two phases of $MoTe_2$, since the Te atoms directly report on structural differences in the two phases. Starting with X-ray absorption (XA) spectroscopy, we resonantly excite from the Te $M_5$ edge into the conduction band of $MoTe_2$, where we then monitor the Te $M_5N_{4,5}N_{4,5}$ Auger decay channel using resonant X-ray photoemission (XP) spectroscopy. The carrier dynamics are extracted from the resulting resonant photoemission (RPE) spectra. Based on the Te $M_5$ XA spectrum and its relation to the conduction band of $MoTe_2$, we determine the underlying mechanism governing the time-evolution of the photoexcited carriers. With our experimental design, we access the sub-femtosecond dynamics near the Weyl point in $T_d$-$MoTe_2$, finding that subtle differences in the electronic structure between the two phases of MoTe$_2$ lead to a carrier lifetime that is a factor 1.5 faster in 1T'-MoTe$_2$ than in T$_d$-MoTe$_2$. We interpret this finding as originating in the decrease in the density of states (DOS) at $E_F$ and a greater screening length resulting from the formation of the Weyl point in T$_d$-MoTe$_2$, ultimately leading to a faster scattering rate of carriers. Our study therefore opens a pathway to understand and potentially monitor and manipulate the fate of carriers in topological quantum materials such as MoTe$_2$.

**Experimental:**

*Sample preparation*

The 1T'-MoTe$_2$ single crystal was purchased from HQ Graphene. The crystal was mounted using double-sided conductive copper tape and cleaved via mechanical exfoliation in an inert argon environment before introduction into the vacuum preparation chamber. To access the T$_d$-phase of MoTe$_2$, the 1T'-MoTe$_2$ crystal was cooled below 110 K using liquid nitrogen. XPS measurements of the Te $3d_{3/2}$ and $3d_{5/2}$ features revealed no oxidation of the MoTe$_2$ crystals.

*Synchrotron measurements*

X-ray absorption and resonant photoemission (RPE) spectroscopy measurements were performed on beamline 10-1 at the Standford Synchrotron Radiation Lightsource (SSRL). Spectra were acquired using a double-pass cylindrical mirror analyzer at an angle of incidence of 55°. The incoming X-rays are linearly polarized in the plane of incidence, exciting both in-plane and out-of-plane orbitals. The XA spectra were measured in total electron yield (TEY) using the sample drain current and the channeltron detector current. The XA and RPE spectra were collected at a pass energy of 50 eV resulting in an energy resolution of ~0.45 eV. The base vacuum pressure during measurements was kept below $5 \times 10^{-9}$ torr.

*Data Treatment*

RPE and XA spectral intensities were initially normalized using the X-ray photon flux ($I_0$) of the incoming soft X-ray radiation measured on a gold grid upstream from the analysis chamber. Due to different sample positions during the RPE measurements, spectral intensities needed to be further normalized. The RPE spectral intensities were normalized using the Te $3d$ features taken at the same photon energies before and after the RPE measurements. Additionally, the sample integrity was checked throughout the RPE measurements using the Te $3d$ features.

RPE spectra were treated with a Shirley background subtraction and XA spectra were treated with a linear background subtraction. The off-resonance RPE data were fit using a Voigt line profile with a Gaussian FWHM of 0.95 eV and a Lorentzian FWHM of 0.65 eV. The Gaussian and Lorentzian FWHM were determined from XP spectra of Au $4f$ and the lifetime of the Te $3d$ core hole, respectively.

**Results:**

As CHC spectroscopy relies on resonant excitation from a core level to an unoccupied band, we first introduce the results of XA spectroscopy. We choose to probe the Te character of the MoTe$_2$ conduction band, exciting from the Te $3d$ core levels, or Te $M_{4,5}$ edge, since this accesses directly regions near the Weyl point (see below). Figure 1a shows the resulting Te $M_{4,5}$ XA spectrum of MoTe$_2$ ranging from 560 to 660 eV. Note that we did not observe major spectral differences between 1T'- and T$_d$-MoTe$_2$ in our XA spectra. This is expected, since the electronic structures differ only subtly, and more sensitive measures such as the electron dynamics are needed to detect the consequences of the structural phase transition. The Te $M_{4,5}$ edge is composed of two small features at 572 eV and 584 eV followed by the so-called "giant resonance" at energies beyond 590 eV.[23] Atomic electric dipole selection rules ($\Delta l = \pm 1$) suggest that the features correspond to the Te($3d$) → Te($5p$) and Te($3d$) → Te($4f$) transitions. The centrifugal barrier for Te core electrons results in a large separation of the Te($5p$) and Te($4f$) states,[23–25] and hence the smaller features correspond to the spin-orbit split Te($3d$) → Te($5p$) transition, while the giant resonance is attributed to Te($3d$) → Te($4f$).

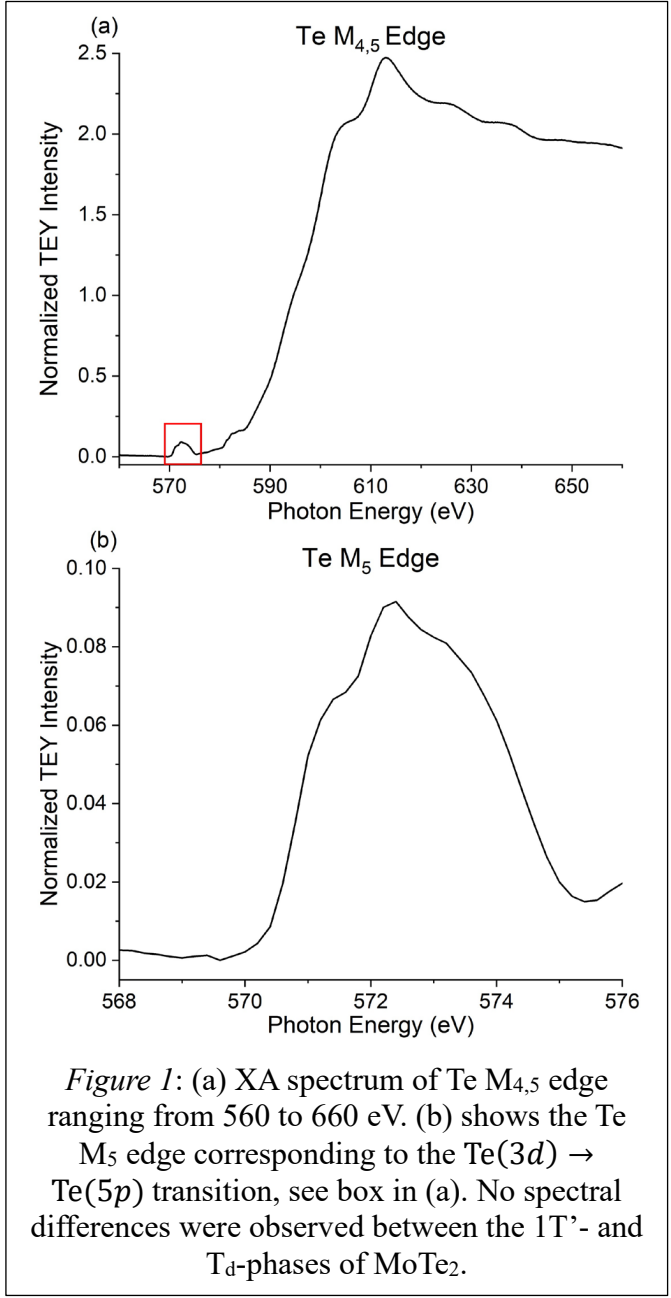

*Figure 1*: (a) XA spectrum of Te $M_{4,5}$ edge ranging from 560 to 660 eV. (b) shows the Te $M_5$ edge corresponding to the Te($3d$) → Te($5p$) transition, see box in (a). No spectral differences were observed between the 1T'- and $T_d$-phases of MoTe$_2$.

We focus on the small features accessing predominately Te $5p$ character at 572 eV and 584 eV, since these constitute selectively transitions to the low-lying topologically relevant states of the conduction bands near $E_F$. In order to simplify the analysis and due to the overlap between the Te $M_4$ edge and the giant resonance, we present here the CHC results of the Te $M_5$ edge only in the remainder of this study. Figure 1b shows a more detailed profile of the Te $M_5$ edge XA spectrum, ranging from the initial onset at 570 eV to 575 eV.

**Normal Auger decomposition and core hole clock spectroscopy**

By sweeping the photon energy over the corresponding energy window of the Te $M_5$ edge and monitoring the Te $M_{4,5}N_{4,5}N_{4,5}$ Auger decay channels in kinetic energy, we map out the RPE spectra for both phases of MoTe$_2$, as shown in Figure 2. Both the 1T'- and T$_d$-phases show resonant enhancement in the Te $M_5N_{4,5}N_{4,5}$ Auger transition at the Te $M_5$ edge, as indicated by the black dashed line. The assignment of this dominant feature in the RPE maps is consistent with the fact that Auger features in the RPE spectra appear at constant kinetic energy, independent of the soft X-ray excitation energy. In contrast, direct photoemission features disperse as a function of photon energy, as is e.g., the case for the Mo 4$s$ and Te 4$p$ features shown in Figure 2 and marked with solid black lines.

To extract the ultrafast dynamics of the conduction band electrons in MoTe$_2$, we take a slice in the RPE map at a specific photon energy and map out the relative intensity of the resonantly excited Auger features. To accomplish this, we use atomic structure theory to model the complete Auger envelope,[26,27] fitting the relative energy splittings and thus the multiplet structure of the Auger spectrum of the Te $M_{4,5}N_{4,5}N_{4,5}$ transition. The Auger envelopes contain contributions from the resonant Auger transition as well as from the corresponding normal Auger transition. The integrated intensity of the resonantly enhanced envelope relative to the normal Auger contribution encodes the element-specific electron dynamics in the conduction band, monitored here on the Te $M_5$ edge of MoTe$_2$. In what follows, we discuss each step in this process in more detail.

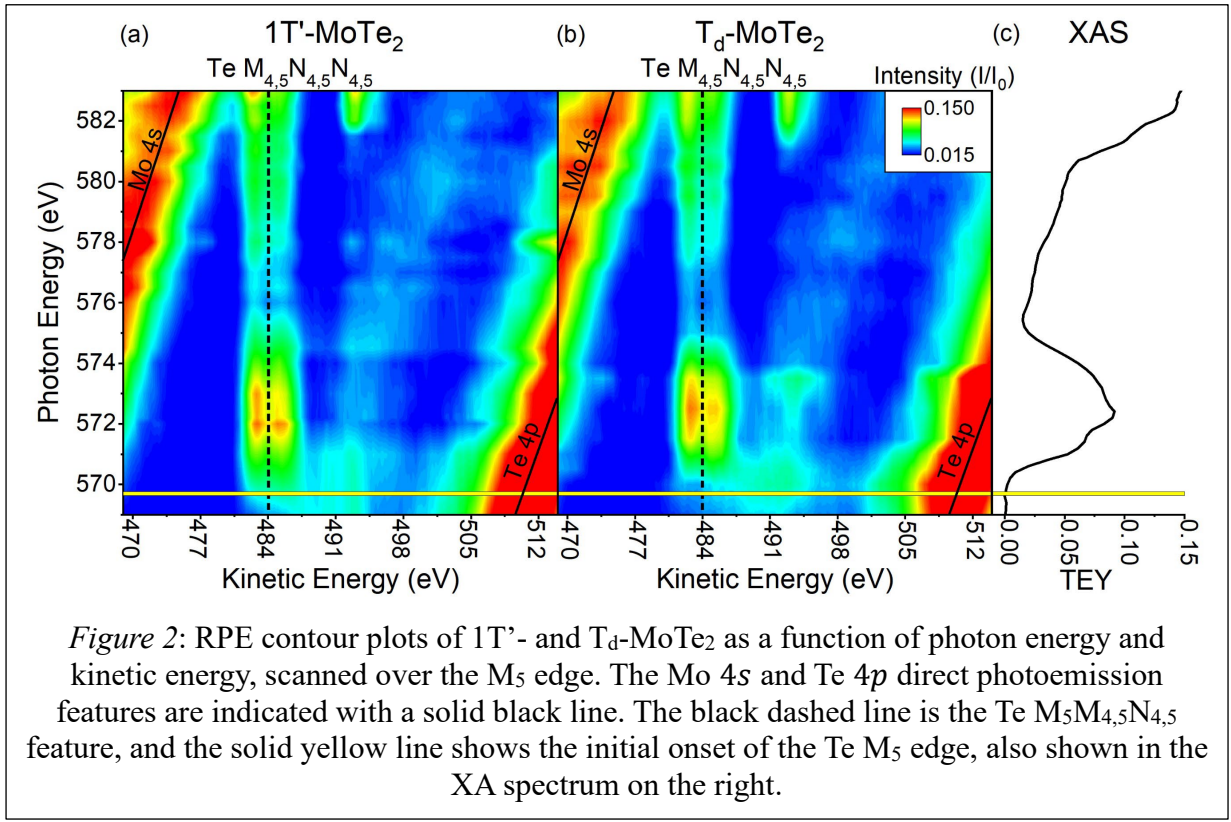

*Figure 2*: RPE contour plots of 1T'- and T$_d$-MoTe$_2$ as a function of photon energy and kinetic energy, scanned over the $M_5$ edge. The Mo 4$s$ and Te 4$p$ direct photoemission features are indicated with a solid black line. The black dashed line is the Te $M_5M_{4,5}N_{4,5}$ feature, and the solid yellow line shows the initial onset of the Te $M_5$ edge, also shown in the XA spectrum on the right.

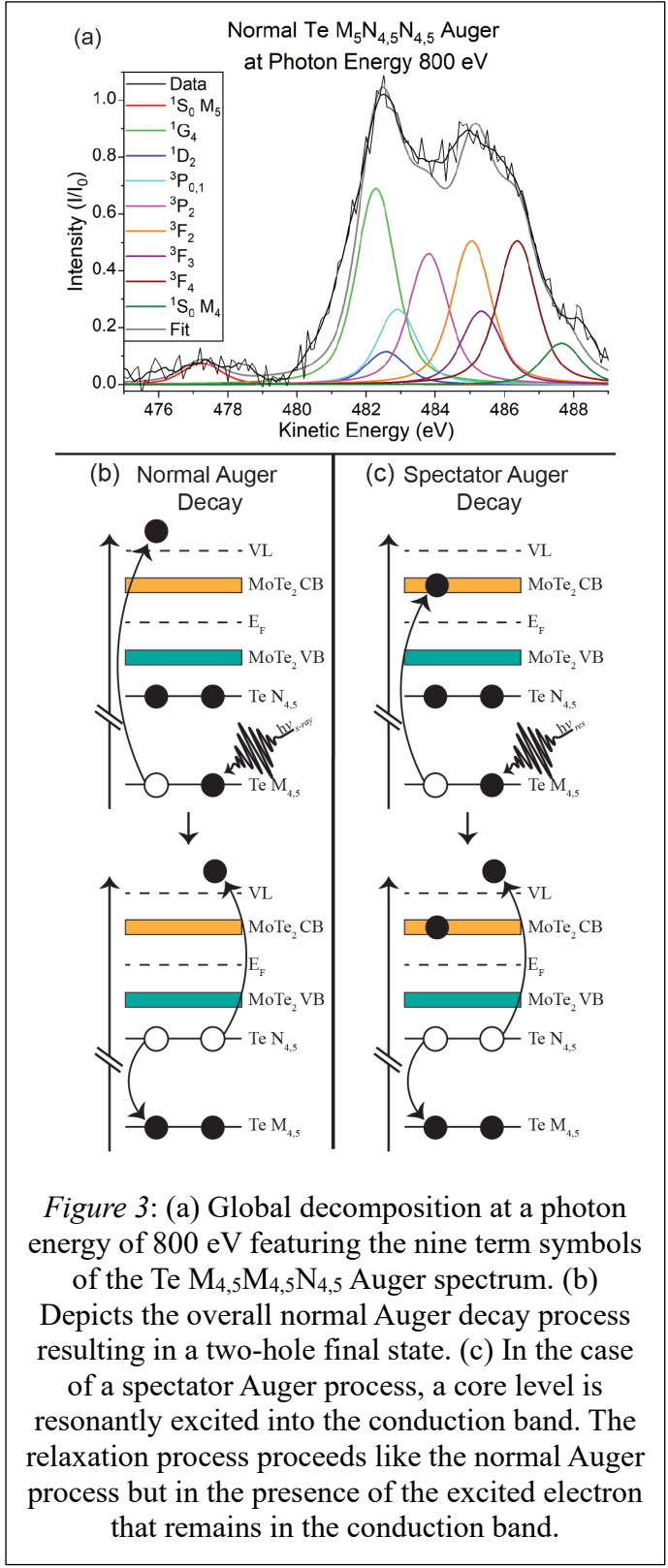

*Figure 3*: (a) Global decomposition at a photon energy of 800 eV featuring the nine term symbols of the Te $M_{4,5}M_{4,5}N_{4,5}$ Auger spectrum. (b) Depicts the overall normal Auger decay process resulting in a two-hole final state. (c) In the case of a spectator Auger process, a core level is resonantly excited into the conduction band. The relaxation process proceeds like the normal Auger process but in the presence of the excited electron that remains in the conduction band.

We begin by modeling the Auger spectral envelope. The Te $M_5N_{4,5}N_{4,5}$ normal Auger transition results in a two-hole final state with a $d^8$ configuration. Consequently, and from atomic structure theory,[26,27] one accesses nine different states with different electronic terms. The energies of these different terms are determined by the Coulombic, exchange, and spin-orbit coupling interactions, and are used to fit the Auger envelope by maximum likelihood estimation **[see SI for more details]**. To exclude contributions from a resonant Auger transition, this fit is carried out far above the Te $M_{4,5}$ edge, where we expect the spectrum to consist principally of the normal (non-resonant) Auger transition. Figure 3a shows the result of this decomposition, encompassing the different terms for the Te $M_5N_{4,5}N_{4,5}$ Auger spectrum. We emphasize that despite significant congestion, these fits are stable and unique since the individual terms and their relative energies are restricted by atomic structure theory. Once the normal Auger envelope is determined, the composite fit containing the nine terms is fixed and used to establish resonant and non-resonant contributions. As shown in Figures 3b and c, the excited electron dynamics of the two phases of $MoTe_2$ are determined based on a competitive Auger decay between the known kinetics of the normal Auger decay channel and the resonant Auger decay channel to be determined.

To understand this in more detail, consider the following competition: During the normal Auger decay process, a core electron is ejected by direct photoemission, forming a core hole with a lifetime on the order of attoseconds to a few femtoseconds. In a many-body decay process, an electron from a higher lying core level fills the core hole, while the excess energy in the system is released by ejecting a second Auger electron, either from the same core level or a higher lying level. The net result is a two-hole final state. In contrast, if the core electron is resonantly excited into the $MoTe_2$ conduction band, the Auger decay process will proceed in a similar way to the normal Auger decay channel but with the photoexcited electron (the spectator electron) residing in the conduction band. The intensity of this channel represents therefore the lifetime of this conduction band electron *relative* to the known normal Auger decay. If the core electron remains localized at the atomic site during the lifetime of the core hole, this resonant Auger channel can be spectroscopically distinguished because the kinetic energy of the outgoing photoelectron is shifted with respect to the normal Auger spectrum: The Coulombic interaction between the two holes which determines the Auger energies is screened by the spectator electron. Hence, the Auger electron carries greater kinetic energy, and the whole Auger envelope is shifted. This is referred to as a spectator shift. Note that the resonant Auger decay process results in a two-hole, one-electron final state. In summary, measuring the ultrafast carrier dynamics reduces to determining the relative intensities of the resonant (or spectator) and normal contributions to the Auger envelope in RPE, as discussed next.

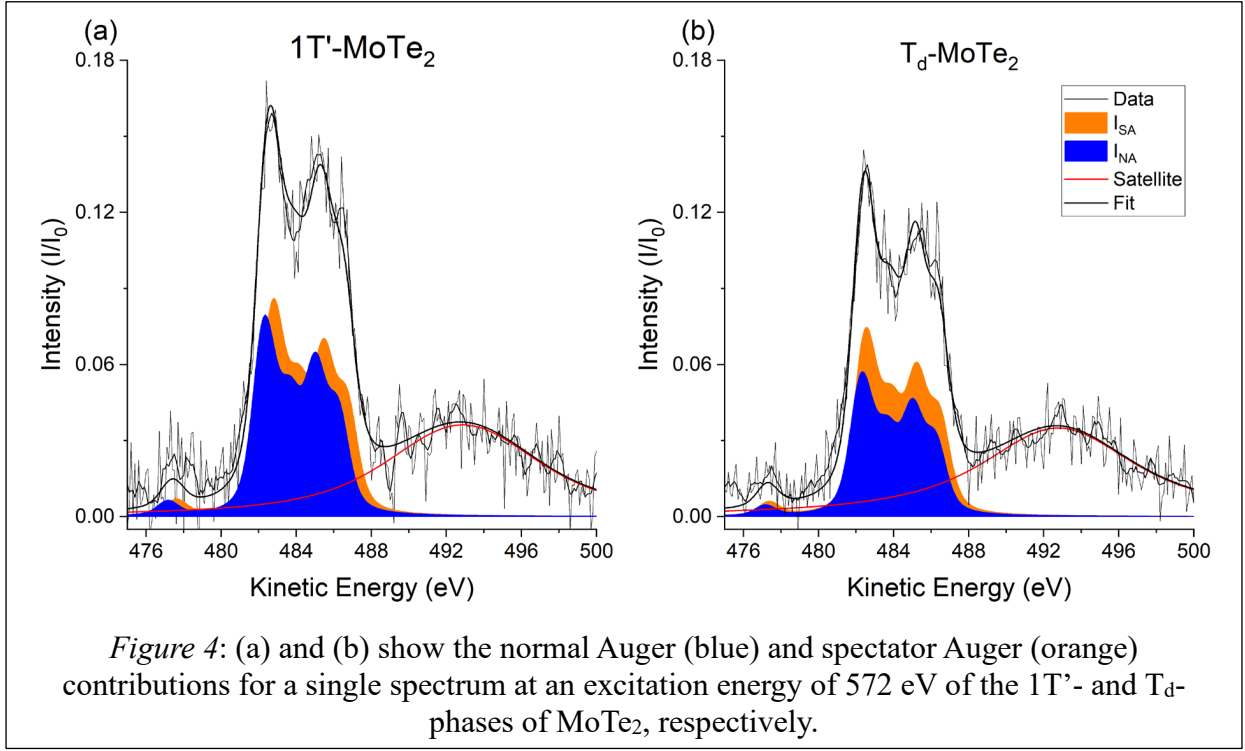

*Figure 4*: (a) and (b) show the normal Auger (blue) and spectator Auger (orange) contributions for a single spectrum at an excitation energy of 572 eV of the 1T'- and $T_d$- phases of MoTe$_2$, respectively.

The resonantly excited Auger features in the RPE maps (Figure 2) contain contributions from both normal and resonant Auger decay processes. We therefore decompose each resonant spectrum across the Te $M_5$ edge with two identical but shifted Auger envelopes, offset with respect to each other by the spectator shift. Over the course of the Te M$_5$ edge, the spectator shifts for the two phases of MoTe$_2$ were optimized and then fixed in a global fit, and the delocalization times are solely dependent on the relative normal and spectator Auger intensity contributions. The respective spectator shifts for the 1T'- and T$_d$-phases are determined to be 0.46 eV and 0.21 eV, respectively. These rather small values are a direct result of the semimetallic nature of MoTe$_2$ which causes strong photohole screening.[19] A representative fit for an excitation energy of 572 eV with the normal (orange) and spectator (blue) Auger contributions is shown in Figures 4a and b for 1T'- and T$_d$-MoTe$_2$, respectively. The broad feature at higher kinetic energies is attributed to a satellite feature.

We extract the relative intensities of the normal and spectator Auger components ($I_{NA}$ and $I_{SA}$) from these fits and plot them as a function of excitation energy. From these intensities, we then extract the carrier delocalization time ($\tau_{del}$) of the resonantly excited electron in the conduction band based on Eqn. 1,

$$\tau_{del} = \tau_{CH} \frac{I_{SA}}{I_{NA}}$$

*Eqn. 1*

where $\tau_{CH}$ is the known lifetime of the Te $3d$ core hole. As is standard in RPE, this lifetime is based on the previously established Lorentzian broadening of high-resolution Auger spectra, and its value is 1.01 fs.[28] Due to the somewhat congested nature of the resonant Auger spectra as a result of the small spectator shifts, we constrain these fits by restricting the normal Auger intensities to vary not more than approximately 10% over the Te M$_5$ resonance. This is justified since normal Auger intensities depend weakly on the initial photoexcitation energy over the range of photon energies used.[18,29]

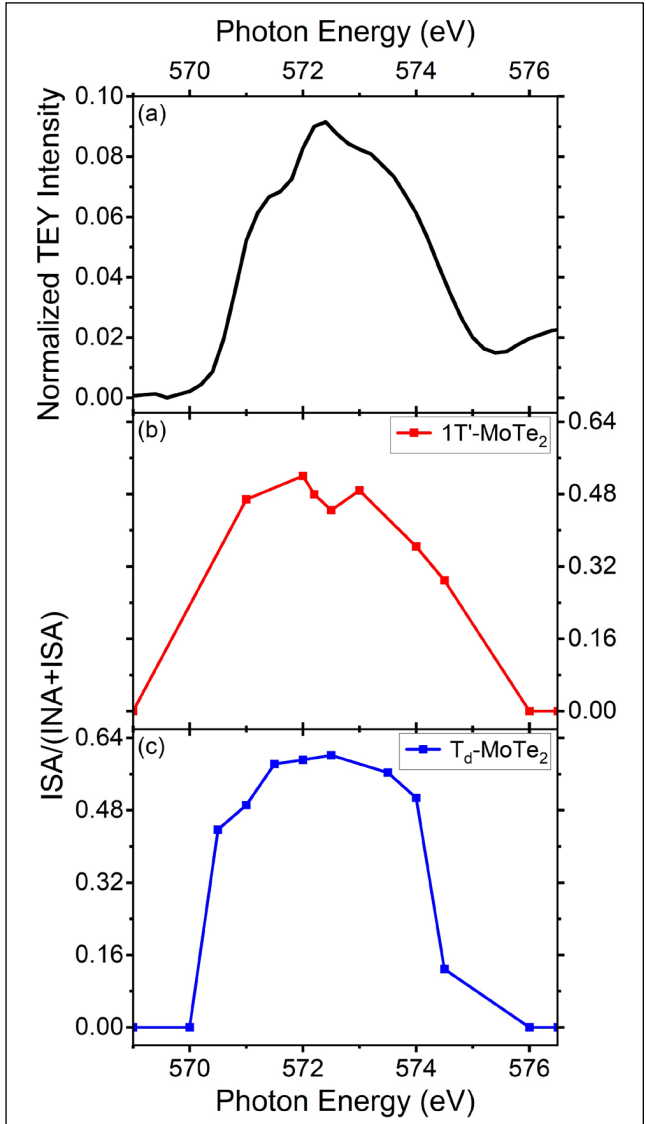

*Figure 5:* Fraction of spectator Auger of the total Auger contribution over the Te $M_5$ edge. (a) XA spectrum, (b) spectator enhancement for 1T'-MoTe$_2$, and (c) spectator enhancement for T$_d$-MoTe$_2$. The enhancement of 1T'- and T$_d$- MoTe$_2$ approximately follows the intensity of the XA spectrum over the $M_5$ edge.

| Average Carrier Delocalization Times of MoTe$_2$ ||
|---|---|
| 1T'-Phase | T$_d$-Phase |
| $950 \pm 100$ as | $1.43 \pm 0.06$ fs |

*Table 1:* Summary of the average carrier delocalization times for the two phases of MoTe$_2$ from 571 to 574 eV.

The results of this analysis are summarized in Figure 5, which shows the XA spectrum and the fraction of spectator Auger of the total Auger contribution as a function of photon energy. Note that in 1T'-MoTe$_2$, Figure 5b, we were not able to include data from photon energies 570 and 570.5 eV. From Eqn. 1, we find carrier delocalization times for the 1T'- and T$_d$- phases of MoTe$_2$ of $950 \pm 100$ as and $1.43 \pm 0.06$ fs, respectively (Table 1). These carrier delocalization times are the average times over an energy window of 571 eV to 574 eV, and the error bars represent the standard deviation over this energy window. The extracted lifetimes lie comfortably within the window accessible to CHC spectroscopy of $\tau_{del} \subset \{0.1\,\tau_{CH}, 10\,\tau_{CH}\}$, which is dictated by the typical signal-to-noise ratio of the soft x-ray beamline at SSRL.

**Discussion:**

The experimentally determined carrier lifetimes of 1T'- and $T_d$-MoTe$_2$ are only accessible because of the extremely fast internal reference "clock" inherent to the CHC spectroscopic approach. Contrary to typical pump-probe spectroscopies and due to the presence of a core hole, the conduction band electrons probed in CHC evolve before hot electron thermalization kicks in. Therefore, CHC spectroscopy affords information that is complementary to the previously reported pump-probe experiments, as discussed next.[14–17]

The power of our CHC approach lies in the fact that it primarily probes the unoccupied density of states (DOS) with Te $5p$ character, as discussed above. Based on DFT+vdW and DFT+U calculations,[30–32] the orbital composition of the electronic structure of 1T'- and $T_d$-MoTe$_2$ near $E_F$ and in the vicinity of Γ is composed of Te $5p$ and Mo $4d$ orbitals, as also corroborated by orbital projected DOS calculations.[31,33] Thus, our RPE experiments directly report on carrier dynamics in bands near points in the Brillouin zone that distinguish the topologically trivial from the WSM phase. To understand the observed differences in carrier delocalization times of 1T'- and $T_d$-MoTe$_2$, we thus consider both the Te orbital contribution to the conduction band of MoTe$_2$ as well as the bonding environment of the Te atoms.

As both phases of MoTe$_2$ are semimetals, the primary difference in the electronic structure is governed by inversion symmetry, preserved in 1T'-MoTe$_2$ and broken in $T_d$-MoTe$_2$. In 1T'-MoTe$_2$, the electron and hole pockets near Γ cross $E_F$ (Figure 6a) and are locally gapped. As a direct result of broken inversion symmetry in the $T_d$ phase, this local gap closes and the electron and hole pockets touch at a single point just above $E_F$ (Figure 6b), constituting a Weyl point.[34,35]

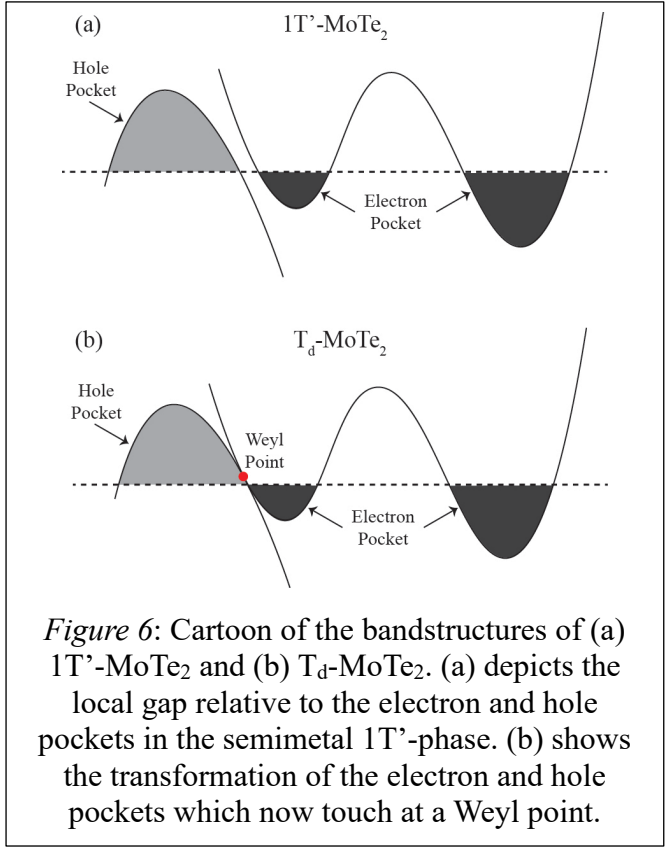

*Figure 6*: Cartoon of the bandstructures of (a) 1T'-MoTe$_2$ and (b) T$_d$-MoTe$_2$. (a) depicts the local gap relative to the electron and hole pockets in the semimetal 1T'-phase. (b) shows the transformation of the electron and hole pockets which now touch at a Weyl point.

We thus examine whether the experimentally determined carrier lifetimes of the two phases are a reflection of the differences in the orbital projected DOS near $E_F$.

The orbital projected DOS over a finite range of energy is directly proportional to the number of states, and hence the number of states into which an excited conduction band electron may scatter. Though the difference is subtle, the closing of the local gap in T$_d$-MoTe$_2$ reduces the DOS near $E_F$ vis-à-vis that of 1T'-MoTe$_2$.[6,33] Hence the scattering rate in the T$_d$ phase is expected to be smaller (longer carrier delocalization time), exactly as observed. Beyond relying on DOS values from electronic structure calculations, optical conductivity measurements confirm this interpretation: These data, directly accessing the near-$E_F$ region, find that as the temperature increases across the phase transition, the chemical potential drops, which again results in an increase of the DOS as the system transitions to the 1T'-MoTe$_2$ phase.[36]

This interpretation is also in agreement with measurements of the DC-dielectric constants as well as STM measurements. The DC-dielectric constants of the two phases are 190.3 and 218.3 for the 1T'-and T$_d$-MoTe$_2$, respectively.[37] As the dielectric constant is inversely proportional to the screening length, a larger dielectric constant reflects strong dielectric screening, thus resulting in reduced scattering in T$_d$-MoTe$_2$. Indeed, STM measurements have shown that impurities increase scattering from the Weyl nodes, lifting and destroying them.[38,39] Thus, Weyl nodes do not act as a scattering center for carriers, indicating that the reduced scattering reflects the longer carrier delocalization time in the WSM $T_d$-MoTe$_2$.

An alternative interpretation may stem from the different structures of the two phases, which result in different bonding environments of the Te atoms. Therefore, we now consider the differences in the crystal structures of 1T'- and $T_d$-MoTe$_2$ as a possible explanation for the different carrier delocalization times observed in CHC. The 1T'- and $T_d$-MoTe$_2$ crystal structures consist of three layers composed of Te-Mo-Te octahedra (see Figure 7). The two crystal structures of MoTe$_2$ differ purely by an interlayer shear displacement which breaks the inversion symmetry in T$_d$-MoTe$_2$. For both crystal structures, the Mo atoms dimerize within a single layer and are shifted off-center from the coordination octahedron, creating a zigzag chain.[13] The deformation of this octahedral environment has two possible orientations which alternate in the unit cell. In the case of T$_d$-MoTe$_2$, this leads to an electrical polarization along the *c*-axis.[40] This constitutes a key structural difference between the 1T'- and T$_d$- phases of MoTe$_2$: For 1T'-MoTe$_2$, the top and bottom layers shift in equal but opposite directions with respect to the middle layer, preserving the inversion symmetry of the unit cell. This shift along the *c*-axis explains the slanted nature of the 1T' unit cell ($\beta = 93.9°$). Conversely, the top and bottom layers in T$_d$-MoTe$_2$ shift in the same direction with respect to the middle layer. This breaks inversion symmetry, shifts the unit cell along the *c*-axis to $\beta = 90°$, and establishes an electric polarization along the *c*-axis.

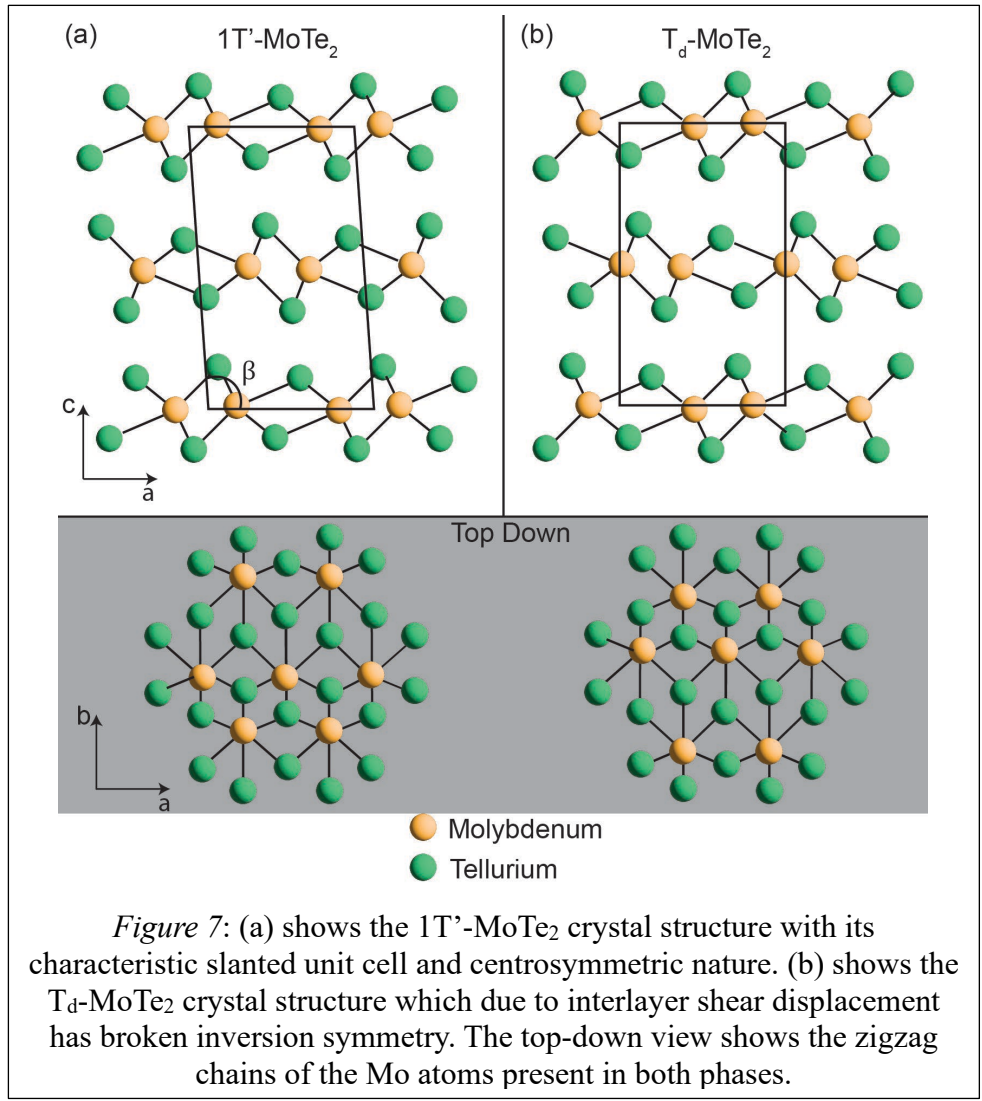

*Figure 7*: (a) shows the 1T'-MoTe$_2$ crystal structure with its characteristic slanted unit cell and centrosymmetric nature. (b) shows the $T_d$-MoTe$_2$ crystal structure which due to interlayer shear displacement has broken inversion symmetry. The top-down view shows the zigzag chains of the Mo atoms present in both phases.

The driving force for the vertical electrical polarization has been attributed to the alternating Te-Te interlayer bond lengths resulting from the interlayer shear displacement in T$_d$-MoTe$_2$.[40] We consider whether a built-in electric field could be responsible for the longer lifetimes observed in T$_d$-MoTe$_2$. The difference in the Te-Te bond lengths is however rather small, and the resulting vertical electrical polarization in T$_d$-MoTe$_2$ is only 0.058 μC cm$^{-2}$.[40] This electrical polarization is significantly smaller than in the structurally related T$_d$-phase of ZrI$_2$ of 0.243 μC cm$^{-2}$.[41] Although the broken inversion symmetry in the T$_d$ phase results in a built-in electric field, we suggest that the changes in the Te bonding environment surrounding the Mo atom are accordingly rather small. Therefore, we do not attribute the factor 1.5 difference in lifetimes of the two phases of MoTe$_2$ to a change in the Te bonding environment.

**Conclusions**

In conclusion, we reveal the carrier delocalization times of the 1T'- and $T_d$-phases of MoTe$_2$ with attosecond time resolution. With our experimental approach, we access the conduction band of MoTe$_2$, probing changes in the carrier dynamics associated with the structural and topological phase transition from 1T'- to $T_d$-MoTe$_2$. Subtle changes in the electronic structure change the screening in $T_d$-MoTe$_2$, thereby decreasing the number of available scattering channels and lengthening the carrier lifetime in the vicinity above $E_F$. Such changes, though not detectable in x-ray absorption spectroscopy, are clearly reflected in the experimentally determined core hole dynamics, highlighting the power of CHC spectroscopy. Understanding the ultrafast carrier dynamics constitutes a step towards harnessing the potential of WSMs for creating robust switchable electronic and spintronic devices.

**Acknowledgments**

We gratefully acknowledge support from the National Science Foundation under Grant No. NSF CHE-1954571. Use of the Stanford Synchrotron Radiation Lightsource, SLAC National Accelerator Laboratory, is supported by the U.S. Department of Energy, Office of Science, Office of Basic Energy Sciences under Contract No. DE-AC02-76SF00515. We also thank Dr. Sami Saino for advice and support.

# Influence of Topology on the Ultrafast Carrier Dynamics in MoTe₂ – Supplementary Information

Sara L. Zachritz[1], Joohyung Park[1], Ayan Batyrkhanov[1], Leah L. Kelly[2], Dennis L. Nordlund[2], and Oliver L.A. Monti[1,3,*]

[1]*Department of Chemistry and Biochemistry, University of Arizona, 1306 East University Boulevard, Tucson, AZ 85721, USA*

[2]*SLAC National Accelerator Laboratory, Stanford Synchrotron Radiation Lightsource, 2575 Sand Hill Road, MS 99, Menlo Park, CA 94025, USA*

[3]*Department of Physics, University of Arizona, 1118 East Fourth Street, Tucson, AZ 85721, USA*

*Corresponding author: monti@arizona.edu

**Normal Auger Decomposition**

The Te $M_5N_{4,5}N_{4,5}$ Auger transitions results in a two-hole final state with $d^8$ electronic configuration. From atomic structure theory, the off-resonant (normal) Auger envelope contains nine different term symbols, $X = {}^1S_0, {}^1D_2, {}^1G_4, {}^3P_{0,1,2},$ and ${}^3F_{2,3,4}$.[26,27] The energy of each term symbol is described in Eqn. 2, which we use to fit the normal Auger envelope:

$$E_{Te_{M_{4,5}N_{4,5}N_{4,5}}}(X) = E(M_{4,5}) - 2E(\overline{N}_{4,5}) - \mathfrak{F}(N_{4,5}N_{4,5}, X) + R_s(N_{4,5})$$

*Eqn. 2*

Here, the first and second terms represent the binding energy of the Te $M_{4,5}$ core level and the average energy of the Te $N_{4,5}$ core level, respectively. The third term, $\mathfrak{F}(N_{4,5}N_{4,5}, X)$, is the energy associated with a specific term symbol, $X$. This term is constructed from Slater integrals, which includes the Coulombic and exchange interactions between the Te $4d$ core holes. Due to the large nuclear mass of the tellurium atom, we also include the spin-orbit coupling interaction to correctly capture the relative energy splittings for the different term symbols. The final term, $R_s$, represent the atomic relaxation which describes the change in screening from gas to solid phase and is estimated based on the values reported by Bahl *et al.*[42]

**Determination of the Spectator Shift**

The spectator shift for the low and room temperature experiments is determined by fitting the resonantly excited Te $M_5N_{4,5}N_{4,5}$ Auger spectra over the Te $M_5$ edge ($hv = 571 - 575$ eV). Using the composite fit of the normal Auger spectrum, we fit each of the resonantly excited Auger spectra using two Auger envelopes representing the intensity contributions from the normal and spectator Auger. For each of the resonantly excited Auger spectra over the Te $M_5$ edge, the energy of the normal Auger envelope is fixed while the energy of the spectator Auger envelope is allowed to float. The final spectator shift is determined based on the global maximum likelihood estimation from these fits over all resonant Auger spectra.